\documentclass[10pt,aps,prb,twocolumn,showpacs,amssymb]{revtex4-1}
\usepackage{amsmath,graphics,epsfig,mathrsfs}
\usepackage[export]{adjustbox}
\usepackage{natbib}
\usepackage{epstopdf}
\usepackage{bm}
\usepackage{color}
\usepackage{braket}
\usepackage{textcomp}
\usepackage{hyperref}
\usepackage{cancel}
\hypersetup{backref=true,pagebackref=true,hyperindex=true,colorlinks=true,breaklinks=true,urlcolor=
blue,linkcolor=blue,bookmarks=true,bookmarksopen=true,citecolor=blue}

\begin{document}
\title{Intrinsic non-adiabatic topological torque in magnetic skyrmions and vortices}
\author{Collins Ashu Akosa}
%\email{collins.akosa@gmail.com}
\author{Papa Birame Ndiaye}
\author{Aur\'elien Manchon}
\email{aurelien.manchon@kaust.edu.sa}
\affiliation{Physical Science and Engineering Division (PSE), King Abdullah University of Science and Technology (KAUST), Thuwal 23955-6900, Kingdom of Saudi Arabia}

\begin{abstract}
We propose that topological spin currents flowing in topologically non-trivial magnetic textures, such as magnetic skyrmions and vortices, produce an intrinsic non-adiabatic torque of the form ${\bf T}_t\sim [(\partial_x{\bf m}\times\partial_y{\bf m})\cdot{\bf m}]\partial_y{\bf m}$. We show that this torque, which is absent in one-dimensional domain walls and/or non-topological textures, is responsible for the enhanced non-adiabaticity parameter observed in magnetic vortices compared to one-dimensional textures. The impact of this torque on the motion of magnetic skyrmions is expected to be crucial, especially to determine their robustness against defects and pinning centers.
\end{abstract}
\maketitle

\section{Introduction}

The search for the efficient electrical manipulation of magnetic textures has recently received a major boost with the observation of ultrafast current-driven domain wall motion in non-centrosymmetric transition metal ferromagnets \cite{Moore2008,Miron2011,Emori2013,Yang2015}, and very low depinning current threshold of bulk magnetic skyrmions \cite{Jonietz2010,Schulz2012,Yu2012}. The latter are topological magnetic defects \cite{Nagaosa2013} that present some similarities with the more traditional magnetic vortices \cite{He2006,Thomas2006,Eltschka2010,Heyne2010}. In both cases, the magnetic topology induces a Lorentz force on the flowing electrons, resulting in topological Hall effect \cite{Neubauer2009,Nagaosa2013} (see also Ref. \onlinecite{Ndiaye2016}). Both classes of magnetic textures also experience a Hall effect when driven by an electric flow, an effect sometimes referred to as skyrmion Hall effect \cite{Nagaosa2013}. In fact, it is important to emphasize that the current-driven characteristics of skyrmions are mostly similar to that of magnetic vortices. For instance, the universal current-velocity relation \cite{Iwasaki2013a,Iwasaki2013b}, as well as the \textit{colossal} spin transfer torque effect at the edge \cite{Iwasaki2014} have already been predicted in the case of magnetic vortices \cite{He2006}. Notwithstanding, skyrmions display some striking differences with respect to magnetic vortices. As a matter of fact, vortex walls are non-local objects composed of a vortex core and two transverse walls \cite{He2006,Thomas2006,Eltschka2010,Heyne2010}, expanding over several exchange lengths in the wire. In contrast, skyrmions are localized objects with a limited expansion from a few tens to hundred nanometers \cite{Nagaosa2013}. This localization has a dramatic impact on the manner skyrmions behave in the presence of pinning centers (defects, notches etc.). When driven by a current, they can deform their texture, and thereby avoid pinning \cite{Iwasaki2013a,Iwasaki2013b,Iwasaki2014,Sampaio2013,Everschor2012,Lin2013}. In contrast, magnetic vortices are much more sensitive to defects and get easily pinned. \par

Since the robustness of skyrmions with respect to defects might hold the key for efficient ultra-dense data storage \cite{Fert2013}, it is crucial to develop a precise understanding of the nature of the torque exerted on these objects. Recently, Iwasaki \textit{et. al.} have shown that such a robustness is partly due to the presence of a large non-adiabatic torque \cite{Iwasaki2013a}. Indeed, in magnetic textures the spin transfer torque can be generally expressed as 
\begin{equation}\label{eq:tor}
{\bf T} = b_{\rm J}({\bf u}\cdot\nabla){\bf M} -\frac{\beta b_{\rm J}}{M_s}{\bf M}\times({\bf u}\cdot\nabla){\bf M},
\end{equation}
where {\bf u} is the direction of current injection, ${\bf m} = {\bf M}/M_s$ is a unit vector in the direction of the magnetization {\bf M} and $M_s$ is the saturation magnetization. The first term in Eq. (\ref{eq:tor}) is the adiabatic torque while the second term is the non-adiabatic torque \cite{Zhang2004,Thiaville2005}. It is well known that while the non-adiabatic parameter in transverse walls is quite small $\beta\approx\alpha$ ($\alpha$ being the magnetic damping of the homogeneous magnet), it is much larger in magnetic vortices, $\beta\approx 8-10\alpha$ \cite{Thomas2006,Eltschka2010,Heyne2010,Bisig2016}. To the best of our knowledge, there is currently no available estimation of the non-adiabaticity parameter in skyrmions, but one can reasonably speculate that it should be of the same order of magnitude as that of magnetic vortices. The recent avalanche of experimental observations of magnetic skyrmions in transition metal multilayers might soon shed light on this question \cite{Jiang2015,Chen2015,Boulle2016,Moreau-Luchaire2016,Woo2016}.\par

That being said, the nature of non-adiabaticity in skyrmions and vortex walls has been scarcely addressed. In transverse domain walls, two major physical mechanisms have been uncovered: spin relaxation \cite{Zhang2004} and spin mistracking \cite{Tatara2004}. Spin relaxation produces a non-adiabatic torque $\beta\sim \tau_{\rm ex}/\tau_{\rm sf}$, where $\tau_{\rm ex}$ is the precession time of the spin about the local magnetization, while $\tau_{\rm sf}$ is the spin relaxation time. Spin mistracking is the quantum misalignement of the flowing electron spin with respect to the local magnetic texture and provides a non-adiabaticity parameter $\beta\sim e^{-\xi \lambda_{\rm dw}}$ that exponentially decreases with the domain wall width $\lambda_{\rm dw}$. It is therefore limited to extremely (atomically) sharp domain walls \cite{Tatara2007,Xiao2006}. We also recently showed that spin diffusion enhances the non-adiabatic torque when the domain wall width is of the order of the spin relaxation length \cite{Akosa2015}. However, none of these effects explains the large difference in non-adiabaticity between transverse walls on the one hand and skyrmions and vortices on the other hand. In a recent work, we proposed that the topological currents induced by the topological Hall effect can enhance the non-adiabatic parameter in magnetic vortices \cite{Bisig2016}. Such an intimate connection between topological Hall effect and non-adiabaticity has also been pointed out by Jonietz \textit{et al.} \cite{Jonietz2010}, but to the best of our knowledge, no theoretical work addresses this issue thoroughly, and all micromagnetic simulations on skyrmions dynamics so far assume a constant $\beta$ parameter \cite{Iwasaki2013a,Iwasaki2013b,Iwasaki2014,Sampaio2013}. \par

As an in-depth follow-up of Ref. \onlinecite{Bisig2016}, the present work investigates the nature of the non-adiabaticity in skyrmions analytically and numerically, and demonstrates that the texture-induced emergent magnetic field inherent in these structures, induces a non-local non-adiabatic torque on the magnetic texture. We provide an explicit expression of the torque and highlight its connection with the spin and charge topological Hall effect.

\section{Emergent Electrodynamics}\label{sec:two}

\subsection{Premises}

It is well-known that when conduction electrons flow in a smooth and slow magnetic texture, ${\bf m}({\bf r} ,t)$, their spin adiabatically changes orientation so that these electrons acquire a Berry phase \cite{Berry1984,Ye1999, Taguchi2001, Onoda2004}. This geometrical phase is attributed to an emergent electromagnetic field $({\bf E}_{\rm em},{\bf B}_{\rm em})$ determined by the magnetic texture gradients \cite{Volovik1987,Bruno2004,Barnes2007,Saslow2007,Tatara2007,Duine2008,Tserkovnyak2008}. Indeed, a time-dependent magnetic texture induces local charge and spin currents through the action of the so-called spin electromotive force \cite{Volovik1987,Barnes2007,Zhang2009,Tanabe2012, Shimada2015,Zang2011,Stern1992}. For the sake of completeness, we derive below this emergent electromagnetic field. Let us considering the simplest Hamiltonian of an \textit{s-d} metal in the presence of a smooth magnetic texture given as 
\begin{equation}\label{eq:ham}
\hat{\cal H}=\frac{\hat{\bf p}^2}{2m} + J_{\rm sd}\hat{\bm\sigma}\cdot{\bf m}({\bf r},t).
\end{equation}
The Schr\"odinger equation corresponding to Eq. (\ref{eq:ham}) can be re-written in the rotating frame of the magnetization, using the unitary transformation $\mathcal{U} = e^{-i\frac{\theta}{2}\boldsymbol{\hat{\sigma}}\cdot{\bf n}}$ where ${\bf n}={\bf z}\times{\bf m}/|{\bf z}\times{\bf m}|$ to obtain 
\begin{equation}
\tilde{\cal H}=\frac{(\hat{\bf p}- e\boldsymbol{\mathcal{A}})^2}{2m} + J_{\rm sd}\hat{\sigma}_z + e\hat{\mathcal{V}},
\end{equation} 
where the vector and scalar potentials are given respectively as $\boldsymbol{\mathcal{ A}}=-\frac{\hbar}{2e} \hat{\bm\sigma}\cdot({\bf m}\times\partial_i{\bf m}){\bf e}_i$ and $\hat{\cal V}=\frac{\hbar}{2e} \hat{\bm\sigma}\cdot({\bf m}\times\partial_t{\bf m})$.
As a consequence, the spin-polarized carriers feel an emerging electromagnetic field on the form \cite{Stern1992,Bruno2004,Barnes2007,Saslow2007,Tatara2007,Duine2008,Tserkovnyak2008,Tanabe2012, Shimada2015,Zhang2009, Zang2011}
\begin{subequations}\label{eq:emf}
\begin{eqnarray}
{\bf E}_{\rm em}^s&=&\frac{s \hbar}{2e} [{\bf m}\cdot(\partial_t{\bf m}\times\partial_i{\bf m})]{\bf e}_i,\\
{\bf B}_{\rm em}^s&=&-\frac{s \hbar}{2e}[{\bf m}\cdot(\partial_x{\bf m}\times\partial_y{\bf m})]{\bf z}.
\end{eqnarray}
 \end{subequations}
The electric field is proportional to the first derivative in time and space and therefore a moving magnetic texture induces a charge current \cite{Barnes2007,Saslow2007,Tatara2007,Duine2008,Tserkovnyak2008} and a self-damping \cite{Zhang2009,Zang2011}. The effect of the magnetic field has so far been overlooked as it requires a second order derivative in space, and is generally considered \textit{small}. Interestingly, this emergent magnetic field turns out to be crucial to understand the spin transport involved in these textures. Indeed,it creates a local ordinary Hall current such that the spin-dependent local charge current driven by the external electric field reads \cite{Bisig2016}
\begin{equation}\label{eq:eff_current}
{\bf j}_e^s =\sigma_0^s {\bf E}+\sigma_0^s{\bf E}_{\rm em}^s+\frac{\sigma_H^s}{B_{\rm H}}{\bf E}\times{\bf B}_{\rm em}^s +\frac{\sigma_H^s}{B_{\rm H}}{\bf E}_{\rm em}^s\times{\bf B}_{\rm em}^s,
 \end{equation}
where $\sigma_0^s$ and $\sigma_H^s$ are respectively the spin-dependent longitudinal and ordinary Hall conductivities, ${\bf E}$ is the external electric field, and $B_{\rm H}=|{\bf B}_{\rm em}^s|$. Inspecting Eq. (\ref{eq:eff_current}), we note that there are two sources of charge or spin currents (i) through the external electric field ${\bf E}$ and (ii) through the emergent electric field driven by the time-variation of the magnetic texture, ${\bf E}_{\rm em}^s$. Therefore, our calculation is able to capture the physics of the motion of the itinerant electrons [Topological (spin) Hall effect] or the magnetic texture itself [skyrmion (vortex) Hall effect]. Let us now assume a rigid magnetic structure for which the time derivative of the magnetization is such that $\partial_t{\bf M} = -({\bf v}\cdot\boldsymbol{\nabla}){\bf M}$, where ${\bf v} = v_x{\bf x} + v_y{\bf y}$ is the velocity of the magnetic structure, and, without loss of generality, an electric field applied along {\bf x} such that ${\bf E} =E{\bf x}$. We can obtain the expressions for the local spin current tensor, ${\mathcal J}_s={\bf M}\otimes({\bf j}_e^\uparrow-{\bf j}_e^\downarrow)$, and charge current vector, ${\bf j}_e={\bf j}_e^\uparrow+{\bf j}_e^\downarrow$, from Eq. (\ref{eq:eff_current}). Explicitly, we obtain
\begin{subequations}\label{eq:currents}
 \begin{eqnarray} \label{eq:currents_a}
\boldsymbol{\mathcal{J}}_s &=&\left[ b_J + \lambda_E^2\left( v_y  + v_x \lambda_H^2\mathcal{N}_{xy}\right)\mathcal{N}_{xy}\right]{\bf M}\otimes{\bf x} \\  \nonumber
&-&\left[\frac{\mathcal{P}_H}{\mathcal{P}_0} \lambda_H^2 b_J + \lambda_E^2  \left(v_x - v_y\lambda_H^2\mathcal{N}_{xy}\right)\right]\mathcal{N}_{xy}{\bf M}\otimes{\bf y},
 \end{eqnarray}  
 \begin{eqnarray} \label{eq:currents_b}
{\bf j}_e &=&\sigma_0\left[E +\frac{ \hbar}{2e} \left(\mathcal{P}_0v_y  + \mathcal{P}_H v_x \lambda_H^2\mathcal{N}_{xy}\right)\mathcal{N}_{xy}\right]{\bf x} \\  \nonumber
&-&\sigma_0\left[\lambda_H^2E + \frac{ \hbar}{2e} \left(\mathcal{P}_0 v_x - \mathcal{P}_Hv_y\lambda_H^2\mathcal{N}_{xy}\right)\right]\mathcal{N}_{xy}{\bf y},
 \end{eqnarray}  
 \end{subequations} 
where $b_J = \hbar \mathcal{P}_{\rm 0}\sigma_0E/2eM_s$, $ \sigma_a = \sigma^{\uparrow}_a + \sigma^{\downarrow}_a$, $\mathcal{P}_a = (\sigma^{\uparrow}_a - \sigma^{\downarrow}_a)/\sigma_a$, (a = 0, {\rm H}), $\lambda_{\rm E}^2 = \hbar^2\sigma_0/4e^2M_s$, $\lambda_{\rm H}^2 = \hbar\sigma_H/2e\sigma_0B_{\rm H}$, and $\mathcal{N}_{\mu\nu}({\bf r}) = {\bf m}\cdot(\partial_\mu{\bf m}\times\partial_\nu{\bf m})$ 
where $\nu, \mu \in (x, y)$. Finally, $\otimes$ is the direct product between spin space and real space. Eqs. (\ref{eq:currents_a}) and (\ref{eq:currents_b}) constitute a central result in this work. They contain information about the motion of itinerant electrons as they traverse a smooth magnetic texture. Indeed, in its explicit form, one is able to explain the subtle difference between non-adiabatic transport in one- and two-dimensional textures. In particular, in addition to the usual constant adiabatic spin current moving along the direction of the applied current ($\sim{\bf x}$) common in one-dimensional textures, the presence of a non-zero topological charge, $\mathcal{N}_{xy} \ne 0$ (as it is the case for magnetic skyrmions and vortices), leads to a texture-induced emergent magnetic field that induces an additional spatially varying spin current along both the longitudinal ($\sim{\bf x}$) and transverse ($\sim{\bf y}$) directions to the electric field ${\bf E}$. This longitudinal spin current is responsible for (i) topological spin and charge Hall effects already observed in topological textures such as skyrmions \cite{Bruno2004,Neubauer2009,Nagaosa2013} (see also Ref. \onlinecite{Ndiaye2016}) and as we propose, (ii) enhanced non-adiabaticity already observed in vortices \cite{Bisig2016}. To clarify a potential mis-conception, we note here that magnetization variation in more than one direction is a \textit{necessary} but not \textit{sufficient} condition to observe these effects. The sufficient condition is a non-zero \textit{topological charge} ($\mathcal{N}_{xy} \ne 0$), which is the case for magnetic textures such as vortices and skyrmions.\par

The topological charge and spin Hall effects arising from the magnetic texture induced-emergent electromagnetic field given by Eq. (\ref{eq:emf}) can be quantified by the charge and spin Hall angles defined respectively as $\theta_{\rm TH}={\rm tan}^{-1}\left(\int j_e^y d^2{\bf r}/ \int j_e^x d^2{\bf r}\right)$ and $\theta_{\rm TSH}={\rm tan}^{-1}\left(\frac{2e}{\hbar}\int \mathcal{J}_s^y d^2{\bf r}/ \int j_e^x d^2{\bf r}\right)$ to obtain
\begin{equation}\label{eq:th}
\theta_{\rm TH} \sim -\mathcal{Q} \lambda_{\rm H}^2 \hspace{2 mm} \mbox{and} \hspace{2 mm}
\theta_{\rm TSH} \sim \mathcal{Q} \mathcal{P}_{\rm H}\lambda_{\rm H}^2,
\end{equation}
where $\mathcal{Q}$ is the topological number defined as $\mathcal{Q} =\frac{1}{4\pi}\int \mathcal{N}_{xy} d^2{\bf r}$.
From Eq. (\ref{eq:th}), one can straightforwardly deduce that $\theta_{\rm TSH} \approx - \mathcal{P}_{\rm H}\theta_{\rm TH}$ which, although very simple, turns out to be far reaching as it captures most of the important physics in static magnetic textures as confirmed by our numerical calculations in the following section.

\subsection{Topological spin torque}

In the previous section, we reminded the basics of the topologically-driven spin and charge currents in magnetic textures. Let us now investigate the impact of this spin current on the dynamics of the magnetic texture itself. By virtue of the spin transfer mechanism, this spin current exerts a torque on the local magnetization, ${\bf T}_{\rm t}=- {\bm\nabla} \cdot {\cal J}_s$, which explicitly reads
\begin{eqnarray}\label{eq:tor_em}
{\bf T}_{\rm t} &=& - \left[b_J + \lambda_{\rm E}^2\left( v_y + v_x\lambda_{\rm H}^2 \mathcal{N}_{xy} \right)\mathcal{N}_{xy}\right]\partial_x{\bf M}\\ \nonumber
&& + \left[\frac{\mathcal{P}_{\rm H}}{ \mathcal{P}_{\rm 0}}\lambda_{\rm H}^2b_J + \lambda_{\rm E}^2\left( v_x - v_y\lambda_{\rm H}^2\mathcal{N}_{xy}\right)\right]\mathcal{N}_{xy} \partial_y{\bf M}. \nonumber
\end{eqnarray}
A remarkable consequence of Eq. (\ref{eq:tor_em}) is that, since $\partial_y{\bf m} \sim {\bf m}\times\partial_x{\bf m}$ in magnetic vortices and skyrmions, the finite topological charge ($\mathcal{N}_{xy} \ne 0$) induces an \textit{intrinsic topological} non-adiabatic spin transfer torque. This non-adiabatic torque is \textit{intrinsic} as it does not rely on impurities or defects (in contrast with the non-adiabaticity studied in Refs. \onlinecite{Zhang2004, Akosa2015}), and \textit{topological} since its origin is associated to the topology of the magnetic texture.  \par

To quantify these effects, we study the dynamics of an isolated magnetic skyrmion and vortex under the action of the torque given in Eq. (\ref{eq:tor_em}) in the context of Thiele formalism of generalized forces acting on a rigid magnetic structure \cite{Thiele1973}. The equation of motion governing the dynamics of these structures is given by the extended LLG equation
\begin{eqnarray}
\partial_t{\bf M}&=&-\gamma{\bf M}\times{\bf H}_{\rm eff} + \frac{\alpha}{M_s}{\bf M}\times\partial_t{\bf M} - {\bf T}
\end{eqnarray} 
where the torque {\bf T} is given as
\begin{equation}\label{eq:tor_tot}
{\bf T} ={\bf T}_{\rm t} + \frac{\beta b_{\rm J}}{M_s}{\bf M}\times\partial_x{\bf M}
\end{equation}
with $\beta$ being a spatially constant non-adiabatic parameter arising from, e.g., spin relaxation \cite{Zhang2004}. To make our analysis simple without missing any interesting physics, we adopt the magnetization profile of an \textit{isolated} skyrmion and vortex core in spherical coordinates as ${\bf m}=(\sin\theta\cos\Phi,\sin\theta\sin\Phi,\cos\theta)$, where the polar angle $\theta$ is defined for an isolated skyrmion as $\cos\theta = p(r_0^2-r^2)/(r_0^2 + r^2)$, and for an isolated vortex as $\cos\theta = p(r_0^2-r^2)/(r_0^2 + r^2)$ for $r \le r_0$, and $\theta = \pi/2 $ for $r >r_0$. $p = \pm 1$ defines the skyrmion (vortex core) \textit{polarity}, $r_0$ defines the radius of the skyrmion (vortex) core. For both textures, the azimutal angle is defined as $\Phi = q Arg(x + iy) + c \pi/2$, where $q = \pm 1$ is the \textit{vorticity} and the $c =\pm 1$ defines the in-plane curling direction otherwise called the \textit{chirality}. For these magnetic profiles, the \textit{topological charge} $\mathcal{Q} =\frac{1}{4\pi}\int \mathcal{N}_{xy} d^2{\bf r}$
is such that $\mathcal{Q} = \frac{1}{2}pq$ for an isolated vortex core and $\mathcal{Q} = (1 - \mathcal{S})pq$ for an isolated skyrmion, where $\mathcal{S} = r_0^2/(r_0^2 +R^2) \to 0$ for $R \gg r_0$. Using these profiles, we obtain the analytical expressions of the velocity components as 
\begin{subequations}\label{eq:velocities}
\begin{eqnarray}\label{eq:vx}
v_x &=& -\frac{\eta_{\rm eff} + \alpha_{\rm eff}\beta_{\rm eff}}{\eta_{\rm eff}^2 + \alpha_{\rm eff}^2}b_J,
\end{eqnarray}
\begin{eqnarray}\label{eq:vy}
v_y &=& pq\frac{\eta_{\rm eff}\beta_{\rm eff} - \alpha_{\rm eff}}{\eta_{\rm eff}^2 + \alpha_{\rm eff}^2}b_J,
\end{eqnarray}
\end{subequations}
where the effective paremeters $\eta_{\rm eff}$, $\beta_{\rm eff}$ and $\alpha_{\rm eff}$ depend on the magnetic texture. For the sake of completeness only, we consider the effect of the renormalization of the gyromagnetic ratio represented by $\eta_{\rm eff}$ which is of order of unity and equals $1 + \frac{16\mathcal{S}_4}{5}\frac{\lambda_{\rm E}^2}{r_0^2}\frac{\lambda_{\rm H}^2}{r_0^2}$ and $1 + \frac{31}{5}\frac{\lambda_{\rm E}^2}{r_0^2}\frac{\lambda_{\rm H}^2}{r_0^2}$ for an isolated skyrmion and vortex respectively, where $\mathcal{S}_k = \sum_{i=0}^{i = k}(\mathcal{S})^i$. However, this is not the focus of this study as this effect is very small and can be neglected.

\subsection{Non-adiabaticity parameter}

The dynamics given by Eqs. (\ref{eq:vx}) and (\ref{eq:vy}) correctly describes the motion of both an isolated skyrmion and vortex core. The dynamics of these two structures is very similar, the major difference being contained in the effective parameters $\eta_{\rm eff}$, $\beta_{\rm eff}$ and $\alpha_{\rm eff}$. The first effective parameter of interest in this study is $\beta_{\rm eff}$, which provides a direct connection between emergent field-induced topological Hall effect and non-adiabaticity. Indeed, while there has been much discussion about the mechanisms responsible for non-adiabatic spin transfer in the literature \cite{Zhang2004, Tatara2004}, the proposed mechanisms although very successful in describing non-adiabatic effects in one-dimensional domain walls, have failed to address the large non-adiabaticity measured in vortex walls. Here, we show that emergent-field induced topological torque gives rise to an additional non-adiabatic torque and a resulting effective non-adiabaticity parameter given as 
\begin{equation}\label{eq:beta}
	 \beta_{\rm eff} = \left\{ \begin{array}{lll}
	  \beta + \frac{4\mathcal{S}_2}{3}\frac{\mathcal{P}_{\rm H}}{ \mathcal{P}_{\rm 0}}\frac{\lambda_{\rm H}^2}{r_0^2} , & \mbox{for skyrmion} \\
      \beta \mathcal{C} + \frac{7}{3}\frac{\mathcal{P}_{\rm H}}{ \mathcal{P}_{\rm 0}}\frac{\lambda_{\rm H}^2}{r_0^2}, & \mbox{for vortex core} , 
	 \end{array}
	 \right.\,
\end{equation}
where the geometric factor $\mathcal{C} = 1 + \ln\sqrt{R/r_0}$. Eq. (\ref{eq:beta}) reveals that associated with these textures is an intrinsic non-adiabaticity parameter which results in an overall enhancement of their one-dimensional and/or non-topological counterpart, $\beta$. This enhancement of the non-adiabaticity parameter is a direct consequence of the topology-induced {\em transverse} spin current ${\cal J}_s^y$, as shown in Eq. (\ref{eq:currents_a}).

\subsection{Damping parameter}

In the context of magnetic textures dynamics, it is practically impossible to discuss the non-adiabaticity parameter $\beta_{\rm eff}$ without a mention of the damping parameter 
$\alpha_{\rm eff}$ as both parameters govern the motion of these structures, see Eq. (\ref{eq:vy}). Different mechanisms for magnetic dissipation have been proposed \cite{Mills2003, Kambersky1976, Kambersky2007,Gilmore2008, Akosa2016a,Zhang2009,Zang2011}. Our calculation reveals that a non-steady state magnetization induces an emergent electric field that results in an intrinsic damping solely due to the topological nature of these textures and thus an overall enhancement of the damping given as
\begin{equation}\label{eq:alpha}
	 \alpha_{\rm eff} = \left\{ \begin{array}{lll}
	 \alpha +\frac{4\mathcal{S}_2}{3} \frac{\lambda_{\rm E}^2}{r_0^2} , & \mbox{for skyrmion}\\
     \alpha \mathcal{C} + \frac{7}{3}\frac{\lambda_{\rm E}^2}{r_0^2}, & \mbox{for vortex}, 
	 \end{array}
	 \right.\,
\end{equation}
which is consistent with Ref. \onlinecite{Zang2011}.

These result are far reaching as they provide a very transparent mechanism which explains the subtle difference in the measured non-adiabaticity in vortices compared one-dimensional domain walls \cite{Bisig2016}. However, as we have shown, two dimensional magnetization variation is a necessary but far from a sufficient condition for this effect, with the sufficient condition being a non-zero topological charge $(\mathcal{N}_{\rm xy} \ne 0)$ as discussed in preceding section.

\section{Numerical Results}

\subsection{Tight-binding model}

\begin{figure}[t!]
\includegraphics[width=8.0cm]{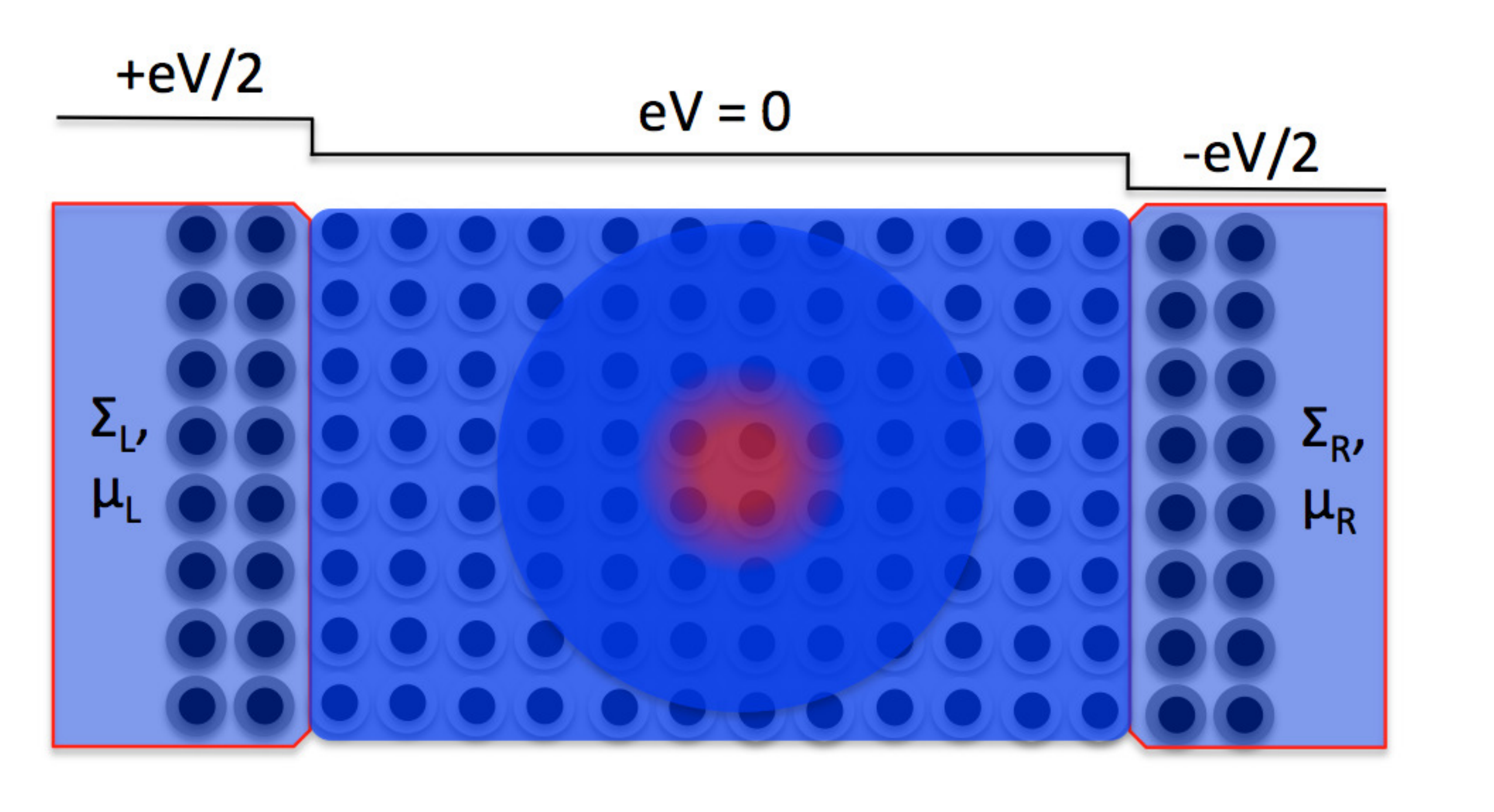}
\caption{Schematic diagram of tight-binding system made up of a central scattering region (isolated magnetic skyrmion/vortex) attached to two ferromagnetic leads (red boxed) L and R at chemical potentials $\mu_L$ and $\mu_R$ respectively. To ensure smooth magnetization variation from the leads to the scattering region, we consider an optimal system size compared to the radius ($R \gg r_0$).}\label{fig:sys_a}
\end{figure}

The derivation proposed in the previous section relied on the adiabatic transport of conduction electrons in a smooth and slowly varying magnetic textures. In other words, the spin of the conduction electrons remains aligned on the local magnetization direction and no spin mistracking is considered \cite{Tatara2004}. In principle, these formulas do not hold when the magnetic texture becomes sharp (i.e. when the skyrmion size is of the order of the spin precession length). In this section, we verify our analytical results numerically using a tight-binding model to test the validity of the adiabatic model discussed in the previous section. The local spin/charge densities and currents as well as the corresponding spin transfer torque are computed numerically using the non-equilibrium wave function formalism \cite{Groth2014}. The system is composed of a scattering region containing an isolated skyrmion or vortex core, attached to two ferromagnetic leads as shown in Fig. \ref{fig:sys_a}. We model this system as a two dimensional square lattice with lattice constant $a_0$, described by the Hamiltonian
\begin{equation} \nonumber
\mathcal{H} = \sum_i \epsilon_i \hat{c}_i^+ \hat{c}_i - t \sum_{<ij>} \hat{c}_i^+ \hat{c}_j - J_{\rm sd} \sum_i \hat{c}_i^+{ \bf m}_i \cdot \boldsymbol{\hat{\sigma}} \hat{c}_i,
\end{equation}
where $\epsilon_i$ is the onsite energy, $t$ is the hopping parameter, the sum $<ij>$ is restricted to nearest neighbors, ${\bf m}_i$ is a unit vector along the local moment at site $i$ coupled by exchange energy of strength $J_{\rm sd}$ to the itinerant electrons with spin represented by the Pauli matrices $\boldsymbol{\hat{\sigma}}$. The label $i $ and $ j$ represent the lattice site and $\hat{c}_i^+=(c_i^\uparrow,c_i^\downarrow)^+$ is the usual fermionic creation operator in the spinor form. 
The $z$-axis is chosen as the quantization axis while the $x$-axis is the direction of current flow. For all our numerical calculations except stated otherwise, we used the parameters: hopping constant $t = 1$, exchange energy $J_{\rm sd} = 2t/3$, onsite energy $\epsilon_i = 0$, transport energy $\epsilon_F = -4.8J_{\rm sd}$, bias $eV = 0.1J_{\rm sd}$, and a large system size of $401\times401a_0^2$ to ensure smooth magnetization variation from system to leads to avoid unphysical oscillations of torques close to the leads. The local spin density ${\bf S}_n^\alpha$ at site $n$ and local spin current density ${\bf J}_{n-1}^\alpha$ between site $n-1$ and $n$ at a particular transport energy for electrons from the $\alpha$-lead can be calculated from their respective operators 
\begin{equation}\label{eq:sden}
{\bf S}_n^{\alpha} = \frac{\hbar}{2}\sum_{\nu} \begin{pmatrix} {\Psi^{\alpha \uparrow}}^{*}_{n\nu}\\ {\Psi^{\alpha \downarrow}}^{*}_{n\nu}\end{pmatrix}^T\boldsymbol{\sigma}\begin{pmatrix}\Psi_{n\nu}^{\alpha \uparrow}\\
\Psi_{n\nu}^{\alpha \downarrow}
\end{pmatrix}
\end{equation}
and 
\begin{equation}\label{eq:curr}
{\bf J}_{n-1}^\alpha = \frac{1}{2i}\sum_{\nu} t_{n\nu, n-1\nu}\begin{pmatrix}{\Psi^{\alpha \uparrow}}^{*}_{n\nu}\\ {\Psi^{\alpha \downarrow}}^{*}_{n\nu}\end{pmatrix}^T\boldsymbol{\sigma}\begin{pmatrix}\Psi_{n-1\nu}^{\alpha \uparrow}\\
\Psi_{n-1\nu}^{\alpha \downarrow}
\end{pmatrix} + c.c,
\end{equation}
where $\Psi_{n\nu}^{\alpha \sigma}$ are $\nu$ propagating mode of the spin-$\sigma$ wave functions from the $\alpha$-lead at site $n$. The quantum mechanical average is calculated by integrating over the small energy window $\epsilon_F - \frac{_{eV}}{^2}$ and $\epsilon_F + \frac{_{eV}}{^2}$ as
\begin{equation}
\langle{\bf S}_n\rangle = \sum_\alpha \int\frac{d\epsilon}{2\pi}f_\alpha {\bf S}_n^{\alpha},
\end{equation}
where $f_\alpha$ is the Fermi-Dirac function for lead-$\alpha$. A similar formula apply for $\langle{\bf J}_{n-1}\rangle$. The corresponding charge density and current density can be obtained from Eqs. (\ref{eq:sden}) and (\ref{eq:curr}) by replacing $\frac{\hbar}{2}\boldsymbol{\sigma}$ by e${\bf I}$ where ${\bf I}$ is the $2\times2$ identity matrix. 

\begin{figure}[t!]
\includegraphics[width=9.0cm]{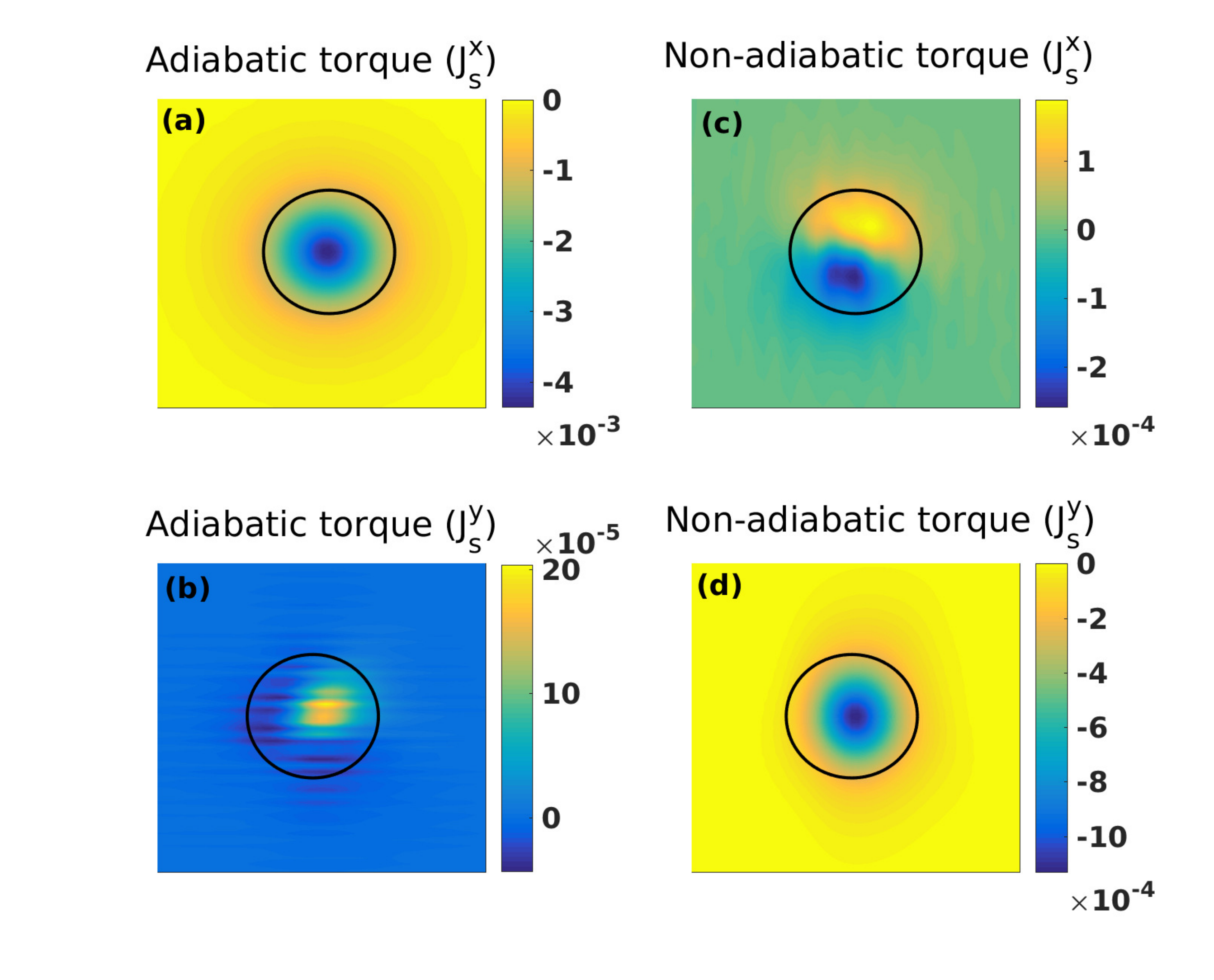}
\caption{Two-dimensional profile of torque components showing contributions from different spin current sources. The adiabatic torque is dominated by the contribution from the longitudinal spin current (a) compared to the transverse spin current (b). The converse is true for the non-adiabatic torque which is dominated by contribution from the transverse spin current (d) compared to the longitudinal spin current (c).}\label{fig:tor2d}
\end{figure}

\subsection{Results and discussion}
\begin{figure}[t!]
\includegraphics[width=6.0cm]{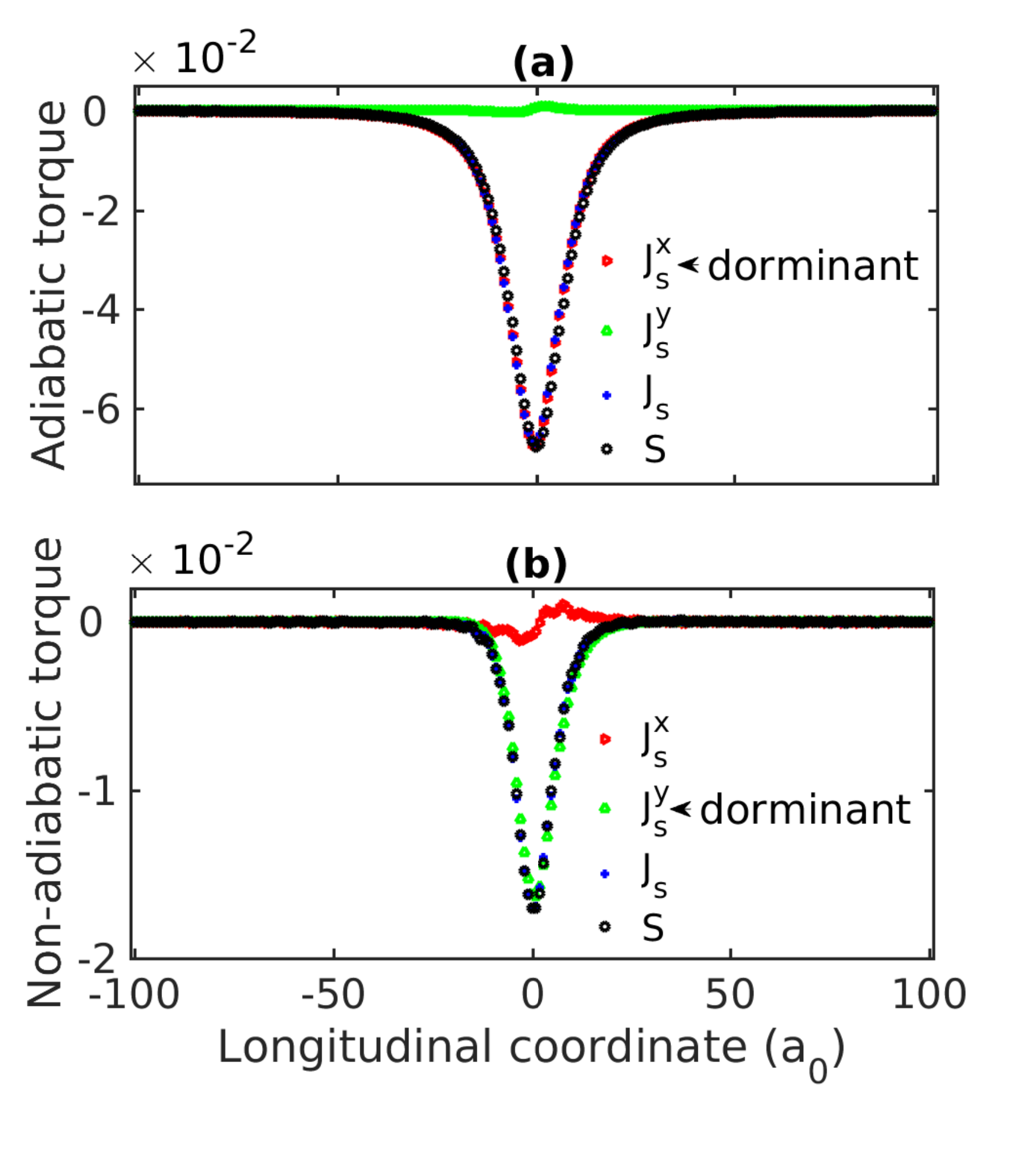}
\caption{One-dimensional profile of torque components showing contribution from different spin current sources and the spin density. The adiabatic torque (a) is dominated by the contribution from $J_s^{x}$ while the non-adiabatic torque (b) is dominated by contribution from $J_s^{y}$. Results shows a very good match between the torque calculated from the spin density ($S$) and that calculated from the spin current $J_s = J_s^{x} + J_s^{y}$.}\label{fig:tor1d}
\end{figure}

Our objective in this section is to ascertain the source of the different contributions to the adiabatic and non-adiabatic torques to uncover their respective microscopic origin and establish a direct correspondence with our theoretical predictions in Section \ref{sec:two}. The most natural and reliable method for calculating the torque {\bf T}$_n$ at site {\bf n} on the local moment {{\bf m}$_n$ is by using the local spin densities as
\begin{equation}\label{eq:torsdn}
{\bf T}_n = \frac{2J_{\rm sd}}{\hbar}\langle{\bf S}_n\rangle \times{\bf m}_n,
\end{equation}
 since this is a conserved quantity, and therefore well defined. However, this method gives little or no information about the microscopic origin of the torque. Therefore, we also calculate the torque from the local spin current as 
 \begin{equation}\label{eq:torcur}
{\bf T}_n = \langle{\bf J}_n\rangle - \langle{\bf J}_{n-1}\rangle. 
\end{equation}
This way, we are able to unambiguously separate the different contributions to the torque arising from the spin current along ${\bf y}$ and ${\bf x}$, quantify its origin, and compare our numerical results with our analytical predictions. A caveat to this method however is that the spin current is a non-conserved quantity especially in systems with spin-orbit coupling or sharp magnetic textures (i.e. sizable spin mistracking). This notwithstanding, since our considerations are based on smooth magnetic textures we can argue that the spin current is well defined and to make sure of this, we used both definitions of the local torque and ensured that the calculated torque using both methods are the same. We deduce the local adiabatic $T_n^{\rm ad}$ and non-adiabatic $T_n^{\rm na}$ torque contribution from the calculated local torque {\bf T}$_n$ by recasting the local torque in the form ${\bf T}_n = T_n^{\rm ad}\partial_x{\bf m}_n - T_n^{\rm na}{\bf m}_n\times\partial_x{\bf m}_n$. Throughout this study except otherwise stated, the torques are reported in units of $2J_{\rm sd}/\hbar$. 

Our numerical results for the two-dimensional profile of spin transfer torque components [calculated using the spin current definition in Eq. (\ref{eq:torcur})] for an isolated skyrmion of core radius $10 a_0$ is shown in Fig. \ref{fig:tor2d}. As shown in Figs. \ref{fig:tor2d} (a) and (b), the adiabatic torque is dominated by the contribution of the {\em longitudinal} spin current ($J_s^{x}$), which is at least an order of magnitude larger than the contribution from the transverse spin current ($J_s^{y}$). The non-adiabatic torque [Figs. \ref{fig:tor2d} (c) and (d)], is largely dominated by contribution of the {\em transverse} spin current ($J_s^{y}$). These results confirm the analysis based on the analytical derivations of the previous section, Eqs. (\ref{eq:currents_a}) and (\ref{eq:tor_em}). \par

Fig. \ref{fig:tor1d} displays the one-dimensional profile of the torque components obtained by summing over the transverse direction for a skyrmion radius $r_0 = 10a_0$. Consistently with Fig. \ref{fig:tor2d}, these results show that the non-adiabatic torque is dominated by the texture-induced {\em transverse} spin current while its adiabatic counterpart is dominated by the {\em longitudinal} spin current. In addition, we also performed these calculations using Eq. ({\ref{eq:torsdn}) (open circles). The results obtained by this method overlap with the results obtained using Eq. (\ref{eq:torcur}). This indicates that for skyrmions with radius $r_0 \ge 10 a_0$ the spin mistracking is negligible.\par

\begin{figure}[t!]
\includegraphics[width=9.5cm]{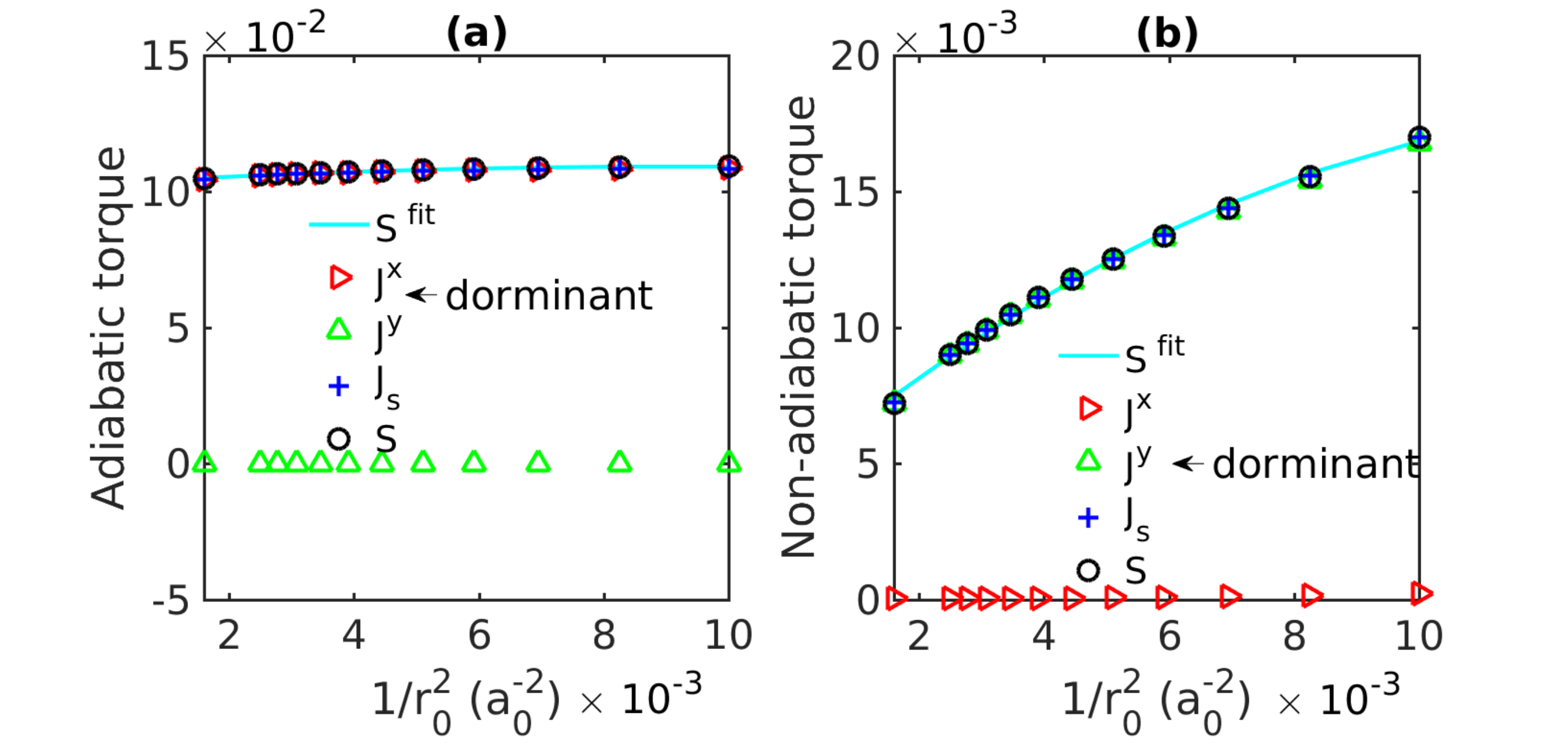}
\caption{Effective adiabatic and non-adiabatic torque dependence on the skyrmion radius. The adiabatic torque (a) is almost non-dependent on the radius of the sykrmion while the non-adiabatic torque (b) shows substantial dependence on the skyrmion radius$r_0$. S$^{\rm fit}$ represents fit of our numerical data to analytical result.}\label{fig:radius}
\end{figure}

Finally, we investigated the scaling laws governing the different contributions of the topological torque components with respect to the skyrmion radius. To achieve this, we calculated the normalized adiabatic and non-adiabatic torque components by integrating the projections of the local torque on ${\bf m}\times\partial_y{\bf m}$ and ${\bf m}\times\partial_x{\bf m}$ respectively and normalize accordingly i.e. 
\begin{subequations}
\begin{equation}
\tilde{{\rm T}}_{\rm ad} = \left[\int {\bf T}\cdot({\bf m}\times\partial_y{\bf m}) d^2{\bf r}\right] /\int M_s \mathcal{N}_{\rm xy} d^2{\bf r}
\end{equation}
\begin{equation}
\tilde{{\rm T}}_{\rm na} = \left[\int{\bf T}\cdot({\bf m}\times\partial_x{\bf m})d^2{\bf r} \right]/\int pqM_s\mathcal{N}_{\rm xy}d^2{\bf r}. 
\end{equation}
\end{subequations}
As shown in Figs. \ref{fig:radius}(a) and (b), while the adiabatic torque is almost independent on the skyrmion radius, the non-adiabatic torque shows a substantial dependence on the skyrmion radius $r_0$ which is in accordance with our analytical results. As a matter of fact, a simple fit of our numerical data to our analytical result given by Eq. (\ref{eq:tor_em}), yields an effective non-adiabatic parameter $\beta_{\rm eff} = 3.5\beta$ for a skyrmion radius of $10a_0$, where $\beta$ is the constant non-adiabaticity parameter obtained in the limit of very large skyrmion radius. For all the range of skyrmion sizes investigated, we also find that the adiabatic assumption exploited to derive the torque expression, Eq. (\ref{eq:tor_em}), is valid. In other words, it confirms that the large non-adiabatic torque in topologically non-trivial magnetic textures can not be explained by spin mistracking.

\section{Conclusion}
We investigated the nature of adiabatic and non-adiabatic spin transfer torque in topologically non-trivial magnetic textures, such as skyrmions and magnetic vortex cores. We showed that the topological spin current flowing through such textures induce an {\em intrinsic topological} non-adiabatic torque, ${\bf T}_t\sim [(\partial_x{\bf m}\times\partial_y{\bf m})\cdot{\bf m}]\partial_y{\bf m}$. Our numerical calculations confirm the physics highlighted by our analytical derivations and confirm that spin transport is mostly adiabatic up to a very good accuracy in these structures, thereby ruling out spin mistracking. Besides providing a reasonable explanation for the enhanced non-adiabaticity in skyrmions and magnetic vortices, our theory opens interesting perspectives for the investigation of current-driven skyrmion dynamics. As a matter of fact, it has been recently proposed that the peculiar robustness of magnetic skyrmions against defects is related to the presence of non-adiabaticity \cite{Iwasaki2013a}. Therefore, understanding the role of this topological non-adiabatic torque when magnetic skyrmions interact with defects is of crucial importance to control current-driven skyrmion motion and achieve fast velocities \cite{Jonietz2010,Schulz2012,Yu2012,Jiang2015,Woo2016}. \par 

CA and AM acknowledges financial support from the King Abdullah University of Science and Technology (KAUST) through the Award No OSR-CRG URF/1/1693-01 from the Office of Sponsored Research (OSR). The authors thank M. Kl\"aui, Kyung-Jin Lee, Gen Tatara, A. Bisig and A. Abbout for inspiring discussions.

%\bibliography{Main_bib.bbl}

\end{document}